\documentclass[onecolumn,aps,prd,superscriptaddress,nofootinbib]{revtex4}
\usepackage{}
\usepackage{graphicx}
\graphicspath{{figures/}{fig/}}
\DeclareGraphicsExtensions{.eps}
\usepackage{amsmath}
\usepackage{bm}
\usepackage{slashed}
\usepackage{epsfig}
\usepackage{amsfonts}
\usepackage{epstopdf}
\usepackage{color}
\usepackage[pdftex]{hyperref}
\usepackage{extarrows}
\usepackage{multirow}
\newcommand{\ben}{\begin{eqnarray}}
\newcommand{\een}{\end{eqnarray}}

\newcommand{\bef}{\begin{figure}[!htp]}
\newcommand{\eef}{\end{figure}}

\newcommand{\bea}{\begin{eqnarray}}
\newcommand{\eea}{\end{eqnarray}}

\def\ba{\begin{linenomath*}\begin{equation}}
\def\ea{\end{equation}\end{linenomath*}}

\allowdisplaybreaks

\newcommand{\state}[4]{{^{#1}\hspace{-0.6mm}#2_{#3}^{[#4]}}}

\newcommand\CScSa{\state{3}{S}{1}{1}}

\newcommand\CScDa{\state{3}{D}{1}{1}}

\newcommand\lrd{\overleftrightarrow{D}}
\newcommand\lrde{\overleftrightarrow{\boldsymbol{\nabla}}}
\newcommand\lrdB{\overleftrightarrow{\nabla}}

\newcommand{\onia}{{\cal Q}}

\newcommand{\sect}[1]{\section{#1}}
\begin{document}
\title{Theory for quarkonium: from NRQCD factorization to soft gluon factorization}

\author{An-Ping Chen}
\email{chenanping@pku.edu.cn}
\affiliation{School of Physics and State Key Laboratory of Nuclear Physics and
	Technology, Peking University, Beijing 100871, China}
\affiliation{Center for High Energy physics, Peking University, Beijing 100871, China}
\author{Yan-Qing Ma}
\email{yqma@pku.edu.cn}
\affiliation{School of Physics and State Key Laboratory of Nuclear Physics and
	Technology, Peking University, Beijing 100871, China}
\affiliation{Center for High Energy physics, Peking University, Beijing 100871, China}
\affiliation{Collaborative Innovation Center of Quantum Matter,
	Beijing 100871, China}

\date{\today}

\begin{abstract}
We demonstrate that the recently proposed soft gluon
factorization (SGF) is equivalent to the nonrelativistic QCD (NRQCD) factorization for heavy quarkonium production or decay, which means that for any given process these two factorization theories are either both valid or both violated. We use two methods to achieve this conclusion. In the first method, we apply the two factorization theories to the physical process $J/\psi \to e^+e^-$. Our explicit calculation shows that both SGF and NRQCD can correctly reproduce low energy physics of full QCD, and thus the two factorizations are equivalent. In the second method, by using equations of motion we successfully deduce SGF from NRQCD effective field theory. By identifying SGF with NRQCD factorization,  we establish relations between the two factorization theories and prove the generalized Gremm-Kapustin relations as a by product. Comparing with the NRQCD factorization, the advantage of SGF is that it resums the series of relativistic corrections originated from kinematic effects to all powers, which gives rise to a better convergence in relativistic expansion.
\end{abstract}

\maketitle
\allowdisplaybreaks

\sect{Introduction}	
\label{sec:intro}

The widely used nonrelativistic QCD (NRQCD) factorization~\cite{Bodwin:1994jh} has encountered some notable difficulties in describing heavy quarkonium data. As the NRQCD factorization is based on the NRQCD effective field theory~\cite{Caswell:1985ui}, it is likely rigorous, although for inclusive quarkonium production only two-loop verification is available at present~\cite{Nayak:2005rt,Nayak:2005rw,Nayak:2006fm}. The main known problem of NRQCD factorization is its bad convergence of relativistic expansion~\cite{Mangano:1996kg}, which may be responsible for its encountered difficulties. Recently, a new factorization approach called soft gluon factorization (SGF)~\cite{Ma:2017xno,Li:2019ncs} was proposed to describe the quarkonium production and decay. The aim of SGF is to resum the series of relativistic corrections originated from kinematic effects in NRQCD, which is the main source to cause the bad convergence in relativistic expansion.

However, the SGF has not been well-established. In this method, hadronization of intermediate quark-antiquark pair to physical quarkonium is described by  nonperturbative soft gluon distributions
(SGDs), which are only formally defined by QCD fields in {\it{small}} loop momentum region~\cite{Ma:2017xno}. Without an explicit definition of {\it{small}} region, it is hard to prove the validity of the SGF for physical processes. Furthermore, the unclear relation between SGF and NRQCD factorization makes it impossible to verify whether kinematic effects have been correctly resummed.

In this paper, with the help of a new regulator, we give a rigorous definition of  {\it{small}} region in SGF. We then provide two strategies to explore the relationship between the SGF and the NRQCD factorization. In the first strategy, we apply the two factorization theories to the physical process of $J/\psi \to e^+e^-$, and show that both the SGF and the NRQCD factorization can correctly reproduce  all the low energy physics of full QCD in this process. In the second strategy, we argue that the SGF formula can be deduced from NRQCD at the operator level by using equations of motion. Both of the two strategies demonstrate that the SGF and the NRQCD factorization are equivalent to each other, which means that for any process these two factorizations theories are either both valid or both violated. By identifying the two theories, we generate complete relations between the nonperturbative matrix element in SGF and NRQCD, which prove the generalized Gremm-Kapustin relations~\cite{Bodwin:2008vp} as a by product.

The rest of this paper is organized as follows. In Sec.~\ref{sec:SGF}, we study the exclusive process $J/\psi \to e^+e^-$ in SGF. We give a rigorous definition of nonperturbative matrix elements in SGF and show that low energy physics of full QCD can be reproduced by SGF.
In Sec.~\ref{sec:NRQCD-SGF}, we discuss the equivalence between SGF and NRQCD factorization and establish relations between the nonperturbative matrix element in SGF and NRQCD. In Sec.~\ref{sec:general}, we show that SGF can be deduced from NRQCD at the operator level. We summarize our results in Sec.~\ref{sec:Conclusion}. Some technical details of our calculation are given in Appendix.~\ref{sec:app}.

\section{$J/\psi \to e^+e^-$ in soft gluon factorization approach }\label{sec:SGF}

\subsection{Factorization formula}

According to Ref.~\cite{Li:2019ncs}, for exclusive decay process, one have the following SGF formula at the amplitude level,
\begin{align}\label{eq:facGA}
\mathcal{A}^{\onia} = \sum_{n} \hat{\mathcal{A} }^{n} \overline{R}^{n*}_{\onia},
\end{align}
where $n$ denote intermediate states. In general, $n$ can contain dynamical soft partons (gluons or light quarks) in
addition to a $Q\bar Q$ pair. But for simplicity, we only discuss intermediate states without dynamical soft partons in this work, although dynamical soft partons can be discussed similarly. Then nonperturbative matrix elements $\overline{R}^{n*}_{{\onia}}$ are defined as
\begin{align}\label{eq:Rt}
\overline{R}^{n*}_{\onia}= \langle 0|[\overline\Psi {\cal K}_{n}\Psi](0) |{\onia}\rangle_S,
\end{align}
where $\Psi$ stands for Dirac field of heavy quark, ${\cal K}_{n}$ is projection
operator defining the intermediate state $n$, and the subscript ``S'' means that,
to evaluate the matrix elements, one only picks up integration
regions where off-shellness of all particles is much
smaller than heavy-quark mass. From the point view of method of regions~\cite{Beneke:1997zp,Jantzen:2011nz}, the effect of
``S'' keeps only {\it small} regions which are everything except the hard region. 

For process $J/\psi \to e^+e^-$, symmetries of QCD tell us that only $n=\CScSa, \CScDa$ are relevant, where we use the spectroscopic notation with superscript ``[1]" denoting color singlet. We thus have
\begin{align}\label{eq:dec1}
\begin{split}
\mathcal{A}^{J/\psi \to e^+e^- }
= \hat{\mathcal{A}}^{\CScSa}\overline{R}^{\CScSa *}_{J/\psi} + \hat{\mathcal{A}}^{\CScDa}\overline{R}^{\CScDa *}_{J/\psi},
\end{split}
\end{align}
with projection operators defined explicitly as
\begin{subequations}\label{eq:Kn}
\begin{align}
{\cal K}_{\CScSa}=& \frac{\sqrt{M}}{M+2m}\frac{M-\slashed{P}}{2M}{\epsilon}_{S_z}^{*\mu} \gamma_\mu\frac{M+\slashed{P}}{2M}{\cal C}^{[1]},\\
\begin{split}
{\cal K}_{\CScDa}=& \frac{\sqrt{M}}{M+2m}\frac{M-\slashed{P}}{2M} {\epsilon}_{S_z}^{*\mu}\gamma^\nu \frac{M+\slashed{P}}{2M}{\cal C}^{[1]}\biggr(-\frac{i}{2}\biggr)^2 \lrd^\alpha\lrd^\beta \biggr( \mathbb{P}_{\alpha\mu}\mathbb{P}_{\beta\nu} -\frac{1}{d-1} \mathbb{P}_{\alpha\beta} \mathbb{P}_{\mu\nu} \biggr) ,
\end{split}
\end{align}
\end{subequations}
where $d=4-2\epsilon$ is the space-time dimension, $D_\mu$ is the gauge covariant derivative with $\overline\Psi \lrd_\mu \Psi =
\overline\Psi (D_\mu \Psi) -
(D_\mu \overline\Psi)\Psi$, $P$ is the momentum of $J/\psi$, ${\epsilon}_{S_z}^{\mu}$ is a polarization vector with $P\cdot {\epsilon}_{S_z}=0 $, $m$ is the heavy-quark mass, $M$ is the mass of $J/\psi$, the color projector is defined as ${\cal C}^{[1]}=1/\sqrt{N_c}$, and the spin projection operator $\mathbb{P}_{\alpha\beta}$ is defined as
\begin{align}
\begin{split}
\mathbb{P}_{\alpha\beta}=-g_{\alpha\beta}+\frac{P_\alpha P_\beta}{P^2}.
\end{split}
\end{align}

The hard parts $\hat{\mathcal{A}}^{n}$ can be perturbatively calculated according to the matching procedure discussed in~\cite{Ma:2017xno,Li:2019ncs}.
To this end, we first replace the $J/\psi$ with a on-shell color-singlet state $c\bar{c}(\CScSa)$ with momenta \footnote{Note that the total momentum of the $c\bar c$ pair $P$ is fixed to the momentum of the physical quarkonium during the matching procedure in SGF.  This is significantly different from the matching procedure in NRQCD factorization, where the total momentum of the pair is a free parameter. }
\begin{align}
p_c&=P/2+q, \quad p_{\overline{c}}=P/2-q
\end{align}
in both sides of Eq.~\eqref{eq:dec1}. On-shell conditions $p_c^2=
p_{\overline{c}}^2 = m^2$ result in
\begin{align}\label{eq:onshel}
P\cdot q =0, \quad \quad q^2=m^2-M^2/4,
\end{align}
which fix $q_0$ and $|\boldsymbol{q}|$ in the rest frame of $P$. The rest of degrees of freedom of $q$ are removed by partial wave expansion, $S$-wave for this case. After the replacement, the l.h.s. of Eq.~\eqref{eq:dec1} becomes
\begin{align}\label{eq:amp-decay}
\begin{split}
\mathcal{A}^{c\bar{c}(\CScSa)\to e^+e^-}
= (-i ee_q)\frac{-i}{M^2}L_{\mu } \int \frac{d^2\Omega}{4\pi} \text{Tr}\left[ \Pi_{1S_z} \mathcal{A}^{\mu }_{c\bar c}\right],
\end{split}
\end{align}
where $\Omega$ is the solid angle of relative momentum $\boldsymbol{q}$ in the $c\bar{c}$ rest frame, $L_\mu$ is the leptonic current
\begin{align}\label{eq:leptonic-current}
L_{\mu }=-ie\,\overline{u}(k_{e^-})\gamma_{\mu} v(k_{e^+}),
\end{align}
and $\mathcal{A}^{\mu }_{c\bar c}$ is the hadronic part of decay amplitude with spinors of $c\bar{c}$ removed. The $c\bar{c}$ pair is projected to state $\CScSa$ by replacing spinors of $c\bar{c}$ pair by
\begin{align}\label{eq:project-operator}
{\Pi}_{1S_z}&=\frac{(\slashed{p}_{{c}}+m) \frac{M+\slashed{P}}{2M} {\epsilon}_{S_z}^{\mu} \gamma_\mu \frac{M-\slashed{P}}{2M} (\slashed{p}_{\overline{c}}-m)}{\sqrt{M}(M/2+m)}{\cal C}^{[1]}.
\end{align}
Similarly we have
\begin{subequations}
	\begin{align}
	\overline{R}^{\CScSa *}_{c\bar{c}(\CScSa)}
	=&  \int \frac{d^2\Omega}{4\pi} \text{Tr}[ \Pi_{1S_z} \overline{R}^{\CScSa *}_{c \bar{c}}],\\
	\begin{split}
	\overline{R}^{\CScDa *}_{c\bar{c}(\CScSa)}
	=&  \int \frac{d^2\Omega}{4\pi} \text{Tr}[ \Pi_{1S_z} \overline{R}^{\CScDa *}_{c \bar{c}}].
	\end{split}
	\end{align}
\end{subequations}
Based on these equations, one can calculate $\mathcal{A}^{c\bar{c}(\CScSa)\to e^+e^-}$, $\overline{R}^{\CScSa *}_{c\bar{c}(\CScSa)}$ and $\overline{R}^{\CScDa *}_{c\bar{c}(\CScSa)}$ perturbatively.

Denoting perturbative expansion of any quantity $W$ as $W=W^{(0)}+\alpha_s W^{(1)}+\alpha_s^2 W^{(2)}+\cdots$, we have the following orthogonal relations~\cite{Ma:2017xno}:
\begin{subequations}
	\begin{align}
	\overline{R}^{\CScSa *,(0)}_{c\bar{c}(\CScSa)}=&1,\\
	\begin{split}
	\overline{R}^{\CScDa *,(0)}_{c\bar{c}(\CScSa)}=&0.
	\end{split}
	\end{align}
\end{subequations}
Based on this, the Eq.~\eqref{eq:dec1} results in the following matching relations:
\begin{subequations}\label{eq:match}
	\begin{align}
	\hat{\mathcal{A}}^{\CScSa,(0)}=&\mathcal{A}^{c\bar{c}(\CScSa) \to e^+e^-,(0) },\\
	\begin{split}
	\hat{\mathcal{A}}^{\CScSa,(1)}=&\mathcal{A}^{c\bar{c}(\CScSa) \to e^+e^-,(1) }-\hat{\mathcal{A}}^{\CScSa,(0)}\overline{R}^{\CScSa *,(1)}_{c\bar{c}(\CScSa)}-\hat{\mathcal{A}}^{\CScDa,(0)}\overline{R}^{\CScDa *,(1)}_{c\bar{c}(\CScSa)},
	\end{split}\\
	\vdots& \nonumber
	\end{align}
\end{subequations}
By replacing the $J/\psi$ by $c\bar{c}(\CScDa)$, we can obtain similar relations for $\hat{\mathcal{A}}^{\CScDa}$. These relations enable us to calculate $\hat{\mathcal{A}}^{\CScSa}$ and $\hat{\mathcal{A}}^{\CScDa}$ perturbatively. For simplicity, we will concentrate on the S-wave contribution $\hat{\mathcal{A}}^{\CScSa}$ in the rest of the paper.

\subsection{Perturbative calculation in full QCD}

We first calculate $\mathcal{A}^{c\bar{c}(\CScSa) \to e^+e^-}$ according to Eq.~\eqref{eq:amp-decay}. The amplitude $\mathcal {A}^\mu_{c\bar c}$ in full QCD can be decomposed as
\begin{align}\label{eq:amplitude}
\begin{split}
\mathcal {A}^\mu_{c\bar c}=\mathcal {G}\gamma^\mu + \mathcal {H}q^\mu ,
\end{split}
\end{align}
and up to order $\alpha_s$ one has~\cite{Bodwin:2008vp},
\begin{subequations}\label{eq:dec-coefficient}
\begin{align}
\mathcal {G} =& 1+ \frac{\alpha_s C_F}{4\pi} \biggr[ 2[ (1+\delta^2)L(\delta) -1 ] \biggr(\frac{1}{\epsilon_{IR}} + \log \frac{4\pi \mu^2 e^{-\gamma_E}}{ m^2 } \biggr) + 6\delta^2L(\delta) -4(1+\delta^2)K(\delta)
\nonumber\\
 &- 4 + (1+\delta^2)\frac{\pi^2}{\delta}\biggr] +\mathcal {O}(\alpha_s^2), \\
\mathcal {H} =& \frac{\alpha_s C_F}{4\pi} \frac{2(1-\delta^2)L(\delta)}{m} +\mathcal {O}(\alpha_s^2)  ,
\end{align}
\end{subequations}
with
\begin{align}
\begin{split}
\delta    =&   \frac{\sqrt{M^2-4m^2}}{M},\\
L(\delta) =&  \frac{1}{2\delta} \log \biggr( \frac{1+\delta}{1-\delta} \biggr),\\
K(\delta) =&  \frac{1}{4\delta} \biggr[ \textrm{Li}_2\biggr(\frac{2\delta}{1+\delta} \biggr)-\textrm{Li}_2\biggr(-\frac{2\delta}{1-\delta} \biggr) \biggr],
\end{split}
\end{align}
and $\textrm{Li}_2$ is the Spence function:
\begin{align}
\begin{split}
\textrm{Li}_2(x)= \int_x^0 dt \frac{\log(1-t)}{t}.
\end{split}
\end{align}
In the above results we have dropped imaginary parts that are irrelevant for our purpose. By inserting Eqs.~\eqref{eq:amplitude} and~\eqref{eq:project-operator} into Eq.~\eqref{eq:amp-decay}, one gets
\begin{align}
\begin{split}\label{eq:amplitude-result}
\mathcal{A}^{c\bar{c}(\CScSa)\to e^+e^-}
= \frac{ee_q}{M^2}  \frac{(2(d-2)M + 4m)\mathcal {G}- (M^2- 4m^2)\mathcal {H}}{(d-1)\sqrt{M}} \sqrt{N_c} {L} \cdot {\epsilon}_{S_z}.
\end{split}
\end{align}
Note that the $M$ in Ref.~\cite{Bodwin:2008vp} is a free parameter. But to use these expressions for SGF, it needs to be the mass of quarkonium.

\subsection{Perturbative calculation of matrix elements in SGF}\label{sec:SGD}

Now we describe our method to calculate $\overline{R}^{\CScSa *}_{c\bar{c}(\CScSa)}$. As was pointed out in Refs.~\cite{Ma:2017xno,Li:2019ncs}, this quantity is defined to include only {\it small} loop momentum region. In the following, we will provide an explicit definition and choose a UV renormalization scheme.

\begin{figure}[htb!]
	\begin{center}
		\includegraphics[width=0.8\textwidth]{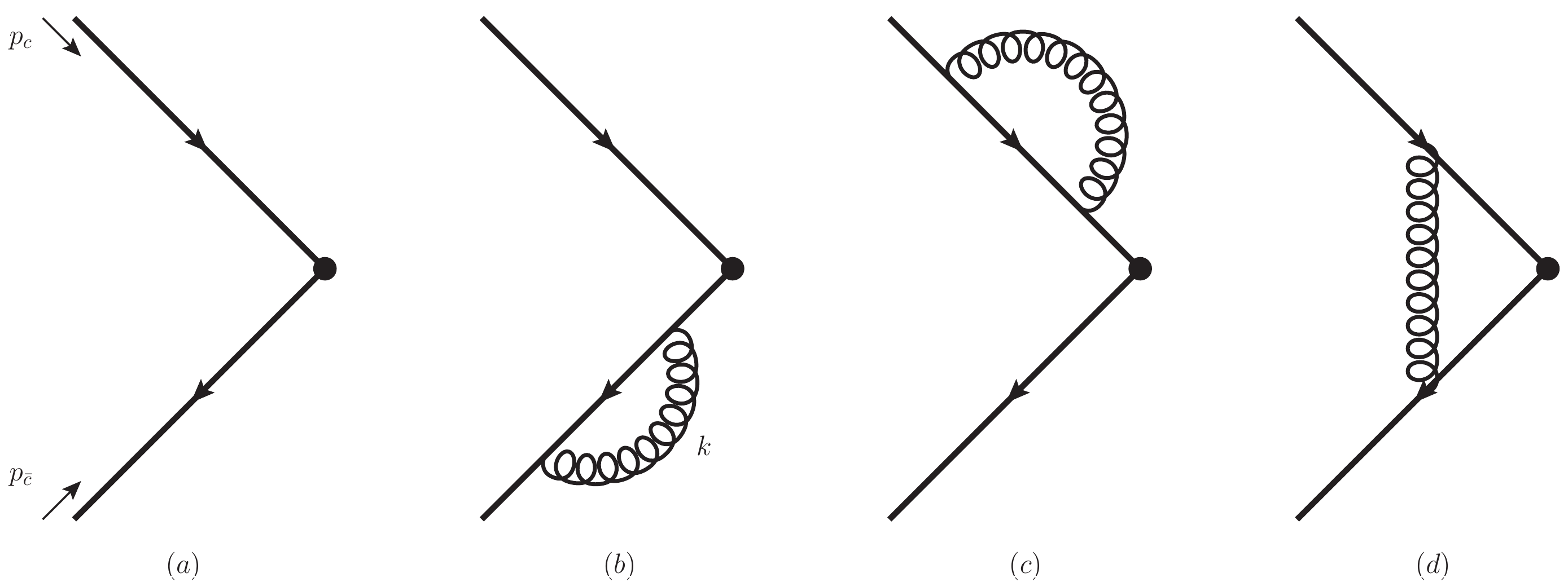}
		\caption{Feynman diagrams for the $\CScSa$ matrix element. \label{fig:diags}}
	\end{center}
\end{figure}

Up to order $\alpha_s$, the corresponding Feynman diagrams are shown in Fig.~\ref{fig:diags}, where the solid circle represents the operator $\overline\Psi {\cal K}_{\CScSa}\Psi$.
The calculation at tree level is straightforward, the result is
\begin{align}
\begin{split}
\overline{R}^{c\bar{c}(\CScSa)*,(0)}_{c\bar{c}(\CScSa)}  =& \int \frac{d^2\Omega}{4\pi} \textrm{Tr} [ {\cal K}_{\CScSa}
\Pi_{1S_z}  ]= 1.
\end{split}
\end{align}

Let's take the vertex correction Fig.~\ref{fig:diags}(d) as an example to explain the calculation at one-loop level. The amplitude reads
\begin{align}
\begin{split}\label{eq:vertex-correction0}
\overline{R}^{c\bar{c}(\CScSa)*}_{c\bar{c}(\CScSa)}\biggr\vert_{d}  =& (-ig_s^2\mu^{2\epsilon})\int \frac{d^2\Omega}{4\pi}\int \frac{d^dk}{(2\pi)^d} T_S\left\{ \frac{\textrm{Tr}  [\gamma^\alpha  (-\slashed{p}_{{\bar c }}+\slashed{k}+m)T^a {\cal K}_{\CScSa} T^a (\slashed{p}_{{ c }}+ \slashed{k} +m)  \gamma_\alpha
	\Pi_{1S_z} ]}{[k^2+i0^+][k^2-2p_{\bar c}\cdot k +i0^+][k^2+2p_{ c}\cdot k +i0^+]}\right\},
\end{split}
\end{align}
where $T_S$ is an operator that forces the loop momentum $k$ to be in {\it small} region~\cite{Ma:2017xno,Li:2019ncs}, which will be defined explicitly by using the method of regions~\cite{Beneke:1997zp,Jantzen:2011nz}.

Before continue, we note that although the full QCD integral in Eq.\eqref{eq:vertex-correction0} is well regularized  by dimensional regularization, the manipulation to define $T_S$ will generate unregularized integrals. Therefore, other regulator is needed to make our manipulation mathematically rigorous. We propose a new regularization at the full QCD level by multiplying the power of all propagator denominators by $1+\eta$, which  can regularize all possible divergences encountered in the derivation using the method of regions, including both ultraviolet and non-ultraviolet divergences \footnote{Non-ultravoilet divergences are necessarily caused by singularities in denominators, which are clearly regularized because we have a power of $\eta$ for all denominators. The power of $\eta$ also makes ultraviolet divergences caused by one or more components of loop momentum going to infinity well regularized. Because effective field theories and factorization theories can be derived from the method of regions, our regularization method is so general that it can regularize all possible divergences in factorization theories and effective field theories. A similar regularization method has been used in the light-cone ordered perturbation theory in Ref.~\cite{Liu:2019iml} to regularize rapidity divergences.}.    For the integral that we are interested in, we get
\begin{align}
\begin{split}\label{eq:vertex-correction}
\overline{R}^{c\bar{c}(\CScSa)*}_{c\bar{c}(\CScSa)}\biggr\vert_{d}  =& (-ig_s^2\mu^{2\epsilon})\int \frac{d^2\Omega}{4\pi}\int \frac{d^dk}{(2\pi)^d} T_S\left\{ \frac{\nu^{3\eta}\textrm{Tr}  [\gamma^\alpha  (-\slashed{p}_{{\bar c }}+\slashed{k}+m)T^a {\cal K}_{\CScSa} T^a (\slashed{p}_{{ c }}+ \slashed{k} +m)  \gamma_\alpha
	\Pi_{1S_z} ]}{[k^2+i0^+]^{1+\eta}[k^2-2p_{\bar c}\cdot k +i0^+]^{1+\eta}[k^2+2p_{ c}\cdot k +i0^+]^{1+\eta}}\right\}+O(\eta),
\end{split}
\end{align}
where $\nu$ is introduced to compensate the mass dimension changed by the new regularization, and we assume $\eta\ll \epsilon$ to make sure that  the theory is eventually regularized by dimensional regularization.

With the new regularization in hands, we decompose the loop momentum $k_\mu=(k_0,\boldsymbol{k})$ into three domains:
\begin{equation}
\begin{array}{l}
{\textrm{the hard domain}: D_h=\{k\in D: k_{0} \gg |\boldsymbol{q}| \vee |\boldsymbol{k}| \gg |\boldsymbol{q}| \}},\\
{\textrm{the soft domain}: D_s=\{k\in D:  |\boldsymbol{k}| \lesssim k_0 \lesssim |\boldsymbol{q}|   \}},\\
{\textrm{the potential domain}: D_p=\{k\in D:  k_0 \ll |\boldsymbol{k}|   \lesssim |\boldsymbol{q}|   \}},
\end{array}
\end{equation}
where relation ``$\lesssim$" is
understood as the negation of ``$\gg$",  $D=\mathbb{R}^{d}$ is the complete integration domain, and an implicit cutoff scale exists to rigorously separate the three domains. This division satisfies
\begin{align}
\begin{split}
&D_i \cap D_j =\emptyset, \quad i,j \in \{h,s,p \}~ \textrm{and}~ i\neq j, \\
&D_h  \cup D_s \cup D_p  =D.
\end{split}
\end{align}
Then for any original integral $F$,
one can split it into the three domains
\begin{align}
\begin{split}\label{eq:split}
F \equiv& \int d^dk I(k,P,q) = \int_{k\in D_h} d^dk I + \int_{k\in D_s} d^dk I +\int_{k\in D_p} d^dk I.
\end{split}
\end{align}
A possible definition of $T_S$ can be $\{k\in D_s\cup D_p\}$, which however involves a hard cutoff that makes it inconvenient to do high order perturbative calculation.

To obtain a more convenient definition, we introduce operators $T^{(i)}~(i \in \{h,s,p \})$ which expand integrand to convergent power series of small quantities in each domain. We also define $T^{(i,j,\cdots)}\equiv T^{(i)}T^{(j,\cdots)}$. Then we have
\begin{align}
\begin{split}\label{eq:exp}
&\int_{k\in D_s} d^dk I +\int_{k\in D_p} d^dk I\\
=&\int_{k\in D_s} d^dk T^{(s)}I +\int_{k\in D_p} d^dk T^{(p)}I\\
=& \int d^dk T^{(s)}I-\int_{k\in D_h} d^dk T^{(s)}I-\int_{k\in D_p} d^dk T^{(s)}I +\int d^dk T^{(p)}I-\int_{k\in D_h} d^dk T^{(p)}I-\int_{k\in D_s} d^dk T^{(p)}I\\
=& \int d^dk T^{(s)}I-\int_{k\in D_h} d^dk T^{(h,s)}I-\int_{k\in D_p} d^dk T^{(p,s)}I +\int d^dk T^{(p)}I-\int_{k\in D_h} d^dk T^{(h,p)}I-\int_{k\in D_s} d^dk T^{(s,p)}I\\
=& \int d^dk \bigg\{T^{(s)}+ T^{(p)}- T^{(s,p)}\bigg\}I-\int_{k\in D_h} d^dk \bigg\{ T^{(h,s)} + T^{(h,p)}- T^{(h,s,p)}\bigg\}I,
\end{split}
\end{align}
where the property $T^{(i,j)}=T^{(j,i)}$ in our case has been used \cite{Jantzen:2011nz}. It is clear that the l.h.s. of the equation and the first term on the r.h.s. of the equation are equivalent in low energy domain, and their difference is an integration in the hard domain which is infrared safe but may be ultraviolet divergent. The difference can be interpreted as a different choice of renormalization scheme. Therefore, we arrive at our final definition:
\begin{align}
\begin{split}\label{eq:TS}
T_S=T^{(s)}+ T^{(p)}- T^{(s,p)},
\end{split}
\end{align}
with UV divergences removed by an $\overline{\textrm{MS}}$ renormalization scheme. $T^{(s)}$ and $T^{(p)}$ are conventionally called soft region and potential region, respectively, while the overlap region $T^{(s,p)}$ removes double counting between $T^{(s)}$ and $T^{(p)}$.

For the soft region contribution $\overline{R}^{c\bar{c}(\CScSa)*,{(s)}}_{c\bar{c}(\CScSa)}$, we apply the operator $T^{(s)}$ to expand the integrand in Eq.\eqref{eq:vertex-correction} for small quantities, which results in the following integrals
\begin{align}
\begin{split}\label{eq:s-expansion0}
\int \frac{dk_0 d^{d-1}\boldsymbol{k}}{(2\pi)^d}  \frac{k_0^{m_1} ( \boldsymbol{k}\cdot \boldsymbol{q})^{m_2}(-k_0^2)^{j_{14}}(\boldsymbol{k}^{2})^{j_{25}}(2 \boldsymbol{k}\cdot \boldsymbol{q})^{j_{36}} }{[k_0^2 - \boldsymbol{k}^{2} + i0^+]^{1+\eta}[ P_0k_0  +i0^+]^{1+\eta+j_{123}}[ - P_0k_0   +i0^+]^{1+\eta+j_{456}}},
\end{split}
\end{align}
where the term $k_0^{m_1} ( \boldsymbol{k}\cdot \boldsymbol{q})^{m_2}$ comes from the numerator in Eq.~\eqref{eq:vertex-correction} and
\begin{align}
\begin{split}
j_{\alpha\beta\cdots} \equiv j_{\alpha} + j_{\beta} +\cdots,
\end{split}
\end{align}
with $j_i$ being non-negative integers.
As scaleless and infrared-finite integrals can be set as zero (this is again a choice of scheme),  we find only integrals with $m_1=m_2=j_1=j_2=j_4=j_5=0$ are relevant,
\begin{align}
\begin{split}\label{eq:s-expansion}
\int \frac{dk_0 d^{d-1}\boldsymbol{k}}{(2\pi)^d}  \frac{(2 \boldsymbol{k}\cdot \boldsymbol{q})^{j_{36}} }{[k_0^2 - \boldsymbol{k}^{2} + i0^+]^{1+\eta}[ P_0k_0  +i0^+]^{1+\eta+j_{3}}[ - P_0k_0   +i0^+]^{1+\eta+j_{6}}}.
\end{split}
\end{align}
The above integrals have pinch poles around $k_0=0$, which can be regularized by the new regulator $\eta$. As integrals in $\overline{R}^{c\bar{c}(\CScSa)*,{(s,p)}}_{c\bar{c}(\CScSa)}$ can be obtained by expanding the $k_0^2$ term in the denominator in Eq.~\eqref{eq:s-expansion}, it is clear that they cancel exactly with the contribution from pinch poles around $k_0=0$ in $\overline{R}^{c\bar{c}(\CScSa)*,{(s)}}_{c\bar{c}(\CScSa)}$. This cancellation shows that the overlap region is important conceptually, although it contains only scaleless integrals that are usually set to zero in dimensional regularization. By combining the soft region and the overlap region, we only need to consider the contribution from the gluon pole and we can take $\eta\to 0$ safely, which results in
\begin{align}
\overline{R}^{\CScSa*,(s)}_{c\bar{c}(\CScSa)}
-\overline{R}^{\CScSa*,(s,p)}_{c\bar{c}(\CScSa)} \biggr\vert_{d}
=&  \frac{\alpha_sC_F}{4\pi}  2(1+\delta^2)L(\delta) \biggr(\frac{1}{\epsilon_{IR}} -\frac{1}{\epsilon_{UV}}  \biggr)    .
\end{align}
We emphasize that, if one wants to calculate $\overline{R}^{\CScSa*,(s)}_{c\bar{c}(\CScSa)}$ separately, the correct order is to do the integration over $k_0$ first with fixed $\boldsymbol{k}$. Or else, poles around $k_0=0$ will be regularized by different regulators between soft region and overlap region which necessarily breaks symmetries of the theory and makes the cancellation between the two regions impossible.

For $\overline{R}^{c\bar{c}(\CScSa)*,{(p)}}_{c\bar{c}(\CScSa)}$,  terms in the expansion are proportional to
\begin{align}
\begin{split}\label{eq:p-expansion}
\int \frac{dk_0 d^{d-1}\boldsymbol{k}}{(2\pi)^d}  \frac{k_0^{m_1} ( \boldsymbol{k}\cdot \boldsymbol{q})^{m_2}(-k_0^2)^{j_{123}} }{[- \boldsymbol{k}^{2} + i0^+]^{1+\eta+j_3}[- \boldsymbol{k}^{2}+ P_0k_0 - 2 \boldsymbol{k}\cdot \boldsymbol{q}  +i0^+]^{1+\eta+j_{1}}[ - \boldsymbol{k}^{2} - P_0k_0 - 2 \boldsymbol{k}\cdot \boldsymbol{q}  +i0^+]^{1+\eta+j_{2}}},
\end{split}
\end{align}
where we can take $\eta\to 0$ as all divergences are well regularized by dimensional regularization. Then because $k_0$ and $\boldsymbol{k}\cdot \boldsymbol{q}$ can be expressed as linear combinations of denominators, if any of $m_1, m_2, j_1, j_2$ and $j_3$ is nonzero the integral can be decomposed to either scaleless and infrared-finite integrals or integrals with pure virtual value. By keeping only real part, we get
\begin{align}
\overline{R}^{\CScSa*,(p)}_{c\bar{c}(\CScSa)}
 \biggr\vert_{d}
=&  \frac{\alpha_s C_F}{4\pi}  (1+\delta^2 )\frac{\pi^2}{\delta}    .
\end{align}

Similarly, for the self-energy diagrams one can derive
\begin{align}
\begin{split}
\overline{R}^{\CScSa*,(s)}_{c\bar{c}(\CScSa)}
-\overline{R}^{\CScSa*,(s,p)}_{c\bar{c}(\CScSa)} \biggr\vert_{b+c}
=&  \frac{\alpha_sC_F}{2\pi}   \biggr( \frac{1}{\epsilon_{UV}} -\frac{1}{\epsilon_{IR}}  \biggr)  ,\\
\overline{R}^{\CScSa*,(p)}_{c\bar{c}(\CScSa)}
 \biggr\vert_{b+c}
=&  0.
\end{split}
\end{align}
Summing these contributions, we obtain the matrix element at NLO before
renormalization
\begin{align}\label{eq:ME-bare}
\begin{split}
\overline{R}^{\CScSa*,(1)}_{c\bar{c}(\CScSa)} \biggr\vert_{\textrm{bare}}=&\overline{R}^{\CScSa*,(s)}_{c\bar{c}(\CScSa)}
-\overline{R}^{\CScSa*,(s,p)}_{c\bar{c}(\CScSa)}
+\overline{R}^{\CScSa*,(p)}_{c\bar{c}(\CScSa)}\biggr\vert_{b+c+d}\\
=&  \frac{\alpha_sC_F}{4\pi} \biggr[    \biggr(\frac{1}{\epsilon_{IR}} -\frac{1}{\epsilon_{UV}}  \biggr) (
2(1+\delta^2)L(\delta)  - 2 ) + (1+ \delta^2) \frac{\pi^2}{\delta} \biggr] .
\end{split}
\end{align}
Ultraviolet divergences in the above result can be removed by
 the $\overline{\textrm{MS}}$ renormalization procedure, which gives renormalized matrix element
\begin{align}\label{eq:ME-NLO}
\begin{split}
\overline{R}^{\CScSa*}_{c\bar{c}(\CScSa)} =&1+ \frac{\alpha_sC_F}{4\pi} \biggr[    \biggr(\frac{1}{\epsilon_{IR}} + \ln(4\pi e^{-\gamma_E})  \biggr) [
2(1+\delta^2)L(\delta)  - 2 ] + (1+ \delta^2) \frac{\pi^2}{\delta} \biggr]  +\mathcal {O}(\alpha_s^2).
\end{split}
\end{align}

Similarly we can find the real part of $\overline{R}^{\CScDa *,(1)}_{c\bar{c}(\CScSa)}$ is proportional to $\overline{R}^{\CScDa *,(0)}_{c\bar{c}(\CScSa)}$, we have
\begin{align}\label{eq:SD}
\overline{R}^{\CScDa *,(1)}_{c\bar{c}(\CScSa)}=0.
\end{align}

\subsection{Matching short-distance hard part up to one-loop order}

Substituting  Eqs.~\eqref{eq:amplitude-result}, \eqref{eq:ME-NLO} and~\eqref{eq:SD} into Eq.~\eqref{eq:match}, we obtain
\begin{subequations}\label{eq:hard-part}
\begin{align}
\hat{\mathcal{A}}^{\CScSa,(0)}
=& \frac{ee_q}{M^2}  \frac{4 (M + m)}{3\sqrt{M}} \sqrt{N_c}  L \cdot {\epsilon}_{S_z} ,\\
\hat{\mathcal{A}}^{\CScSa,(1)}
= & \frac{ee_q}{M^2}  \frac{4 (M + m)\mathcal {G}^\prime- (M^2- 4m^2)\mathcal {H}^\prime}{3\sqrt{M}} \sqrt{N_c}{L} \cdot {\epsilon}_{S_z},
\end{align}
\end{subequations}
where
\begin{subequations}\label{eq:HP-function}
\begin{align}
\mathcal {G}^\prime =&  \frac{\alpha_s C_F}{4\pi} \biggr[ 2( (1+\delta^2)L(\delta) -1 ) \log \biggr( \frac{\mu^2 }{ m^2 } \biggr)  + 6\delta^2 L(\delta) -4(1+\delta^2)K(\delta)
- 4 \biggr], \\
\mathcal {H}^\prime =& \frac{\alpha_s C_F}{4\pi} \frac{2(1-\delta^2)L(\delta)}{m}   .
\end{align}
\end{subequations}
We find the matrix element defined in SGF reproduces all infrared and Coulomb divergences in full QCD, and the obtained hard part is free of divergences. Therefore we conclude that the SGF factorization holds at least at one-loop level.

\subsection{Validity of SGF at all orders in $\alpha_s$}

The correctness of SGF at one-loop order can be understood in the following way. The full QCD results in Eq.~\eqref{eq:amplitude-result} can also be reproduced by the method of regions (see   Appendix \ref{sec:app} for details), in which $\mathcal{A}^{c\bar{c}(\CScSa)\to e^+e^-}$ is expressed as
\begin{align}
\begin{split}
\mathcal{A}^{c\bar{c}(\CScSa)\to e^+e^-,(1)}
=& \bigg\{\mathcal {A}^{(h)}
-\mathcal {A}^{(h,s)}-\mathcal {A}^{(h,p)}
+\mathcal {A}^{(h,s,p)} \bigg\}   + \bigg\{ \mathcal {A}^{(s)}+ \mathcal {A}^{(p)}-\mathcal {A}^{(s,p)} \bigg\}.
\end{split}
\end{align}
Where the first term on r.h.s has nonzero support only in the hard domain and thus is infrared safe. It is straight forward to check that the low energy part of $\mathcal{A}^{c\bar{c}(\CScSa)\to e^+e^-}$ has been correctly reproduced by matrix element in SGF, i.e.,
\begin{align}
\begin{split}
\mathcal {A}^{(s)}+ \mathcal {A}^{(p)}-\mathcal {A}^{(s,p)} = \hat{\mathcal{A}}^{\CScSa,(0)} \overline{R}^{\CScSa*,(1)}_{c\bar{c}(\CScSa)} \biggr\vert_{\textrm{bare}},
\end{split}
\end{align}
which leaves the corresponding short-distance hard part defined by high energy part of  $\mathcal{A}^{c\bar{c}(\CScSa)\to e^+e^-}$, and thus infrared safe. More precisely, we have
\begin{align}
\begin{split}
\hat{\mathcal{A}}^{\CScSa,(1)}
=& \bigg\{\mathcal {A}^{(h)}
-\mathcal {A}^{(h,s)}-\mathcal {A}^{(h,p)}
+\mathcal {A}^{(h,s,p)} \bigg\}\bigg|_{\overline{\textrm{MS}}},
\end{split}
\end{align}
where $\overline{\textrm{MS}}$ means to remove UV divergences by $\overline{\textrm{MS}}$ subtraction scheme\footnote{Because overlap regions are scaleless which can be set to zero in dimensional regularization, a simpler way to obtain short-distance hard part is to use $\mathcal {A}^{(h)}|_{\overline{\textrm{MS}}}$. But keep in mind that $\overline{\textrm{MS}}$ means to remove IR divergences by $\overline{\textrm{MS}}$ subtraction scheme.}.

The above one-loop argument can be generalized to all orders. By definition, low energy part (``small" region of loop momenta) in full QCD can be reproduced by matrix element in SGF at any order in $\alpha_s$ expansion, with a proper definition of $T_S$ at multi-loop level as discussed in Ref.~\cite{Jantzen:2011nz}. Therefore, the short-distance hard part is perturbatively infrared-safe, which means the SGF formula for the decay width of $J/\psi \to e^+e^-$ holds to all orders in $\alpha_s$.

\section{Relation between SGF and NRQCD factorization for $J/\psi \to e^+e^-$}\label{sec:NRQCD-SGF}

\subsection{NRQCD result}

Ignoring operators involving gauge fields, the NRQCD factorization for the decay amplitude $J/\psi \to e^+e^-$ is given by ~\cite{Bodwin:2007fz,Lee:2018aoz,Bodwin:2008vp}
\begin{align}
\begin{split}\label{eq:NRQCD-form}
\mathcal{A}^{J/\psi \to e^+e^- }
=&  \sum_{n} s_n(\CScSa) \langle 0 \vert \mathcal{O}_{An} \vert J/\psi \rangle + \sum_{n} b_n(\CScDa) \langle 0 \vert \mathcal{O}_{Dn} \vert J/\psi \rangle,
\end{split}
\end{align}
with \footnote{Note that here the normal derivative is used instead of gauge covariant derivative. They are equivalent for our discussion because operators involving gauge fields are ignored.}
\begin{subequations}
\begin{align}
\langle 0 \vert\mathcal{O}_{An} \vert J/\psi \rangle =&  \langle 0 | \chi^\dag \biggr(-\frac{i}{2}\lrde \biggr)^{2n}   \boldsymbol{\sigma} \cdot \boldsymbol{\epsilon}_{S_z}^{*} \psi | J/\psi \rangle ,\\
\langle 0 \vert\mathcal{O}_{Dn} \vert J/\psi \rangle =&  \langle 0 | \chi^\dag \biggr(-\frac{i}{2}\lrde \biggr)^{2n-2}  \biggr[  \biggr(-\frac{i}{2} \lrde \cdot \boldsymbol{\epsilon}_{S_z}^{*}  \biggr) \biggr(-\frac{i}{2} \lrde \cdot \boldsymbol{\sigma} \biggr) - \frac{1}{d-1} \biggr(-\frac{i}{2} \lrde \biggr)^2 \boldsymbol{\sigma}\cdot \boldsymbol{\epsilon}_{S_z}^{*} \biggr]\psi | J/\psi \rangle,
\end{align}
\end{subequations}
where $\psi$ and $\chi^\dag$ are the two-component heavy quark fields in NRQCD, $\lrdB$ is defined as $\chi^\dagger \lrdB \psi =
\chi^\dagger (\nabla \psi) -
(\nabla\chi^\dagger)\psi$, and  $s_n(\CScSa)$ and $ b_n(\CScDa)$ are short-distance hard parts which can be perturbatively calculated. The first two orders in $\alpha_s$ expansion for $s_n(\CScSa)$ are given by~\cite{Bodwin:2007fz,Lee:2018aoz,Bodwin:2008vp}
\begin{subequations}\label{eq:NRQCD-SDC}
\begin{align}
s_n^{(0)}(\CScSa)
=& - ee_q    {L} \cdot {\epsilon}_{S_z} \biggr[\frac{1}{n!} \biggr(\frac{\partial}{\partial \boldsymbol{q}^2} \biggr)^{n} \biggr(\frac{2\sqrt{2}(M + m)}{3M^2\sqrt{M}}\biggr) \biggr] \biggr\vert_{\boldsymbol{q}^2=0},\\
s_n^{(1)}(\CScSa)
=& - ee_q    {L} \cdot {\epsilon}_{S_z}\biggr[\frac{1}{n!} \biggr(\frac{\partial}{\partial \boldsymbol{q}^2} \biggr)^{n} \biggr(\frac{2\sqrt{2}(M + m)\mathcal {G}^\prime - 2\sqrt{2}\boldsymbol{q}^2 \mathcal{H}^\prime}{3 M^2\sqrt{M}} \biggr) \biggr] \biggr\vert_{\boldsymbol{q}^2=0},
\end{align}
\end{subequations}
with $\mathcal{G}^\prime$, $\mathcal{H}^\prime$ given in Eq.~\eqref{eq:HP-function}.

In Ref.~\cite{Bodwin:2008vp}, the S-wave contributions in Eqs.~\eqref{eq:NRQCD-form} and~\eqref{eq:NRQCD-SDC} are further resummed
by using the generalized Gremm-Kapustin relation~\cite{Bodwin:2008vp,Bodwin:2006dn,Gremm:1997dq}
\begin{align}
\begin{split}\label{eq:general-Gremm-Kapustin}
\langle \boldsymbol{q}^{2n} \rangle_{J/\psi} = \langle \boldsymbol{q}^{2} \rangle_{J/\psi}^{n},
\end{split}
\end{align}
with
\begin{align}
\begin{split}
\langle \boldsymbol{q}^{2n} \rangle_{J/\psi} \equiv& \frac{\langle 0 \vert\mathcal{O}_{An} \vert J/\psi \rangle}{\langle 0 \vert\mathcal{O}_{A0} \vert J/\psi \rangle},\\
\langle \boldsymbol{q}^{2} \rangle_{J/\psi} =& m(M-2m)(1 + \mathcal {O}(v^2)).
\end{split}
\end{align}
These relations are obtained by computing the matrix element $\langle 0 \vert\mathcal{O}_{An} \vert J/\psi \rangle$ in potential-model~\cite{Bodwin:2006dn}.

\subsection{Equivalence between SGF and NRQCD factorization}

The basic reason for the validity of SGF in this process is that SGF matrix elements reproduce low energy part of full QCD. As NRQCD matrix elements also correctly reproduce low energy part of full QCD, the SGF is equivalent to NRQCD factorization. It means that, for any process, the two factorization formulas are either both valid or both broken.

Because the amplitude of $J/\psi \to e^+e^-$ can be factorized in both SGF and NRQCD factorization, we have ($D$-wave contributions are suppressed)
\begin{align}
\begin{split}
\mathcal{A}^{J/\psi \to e^+e^- }= \hat{\mathcal{A}}^{\CScSa}\overline{R}^{\CScSa *}_{J/\psi}(1 + \mathcal {O}(v^2))
=&  \sum_{n} s_n(\CScSa) \langle 0 \vert \mathcal{O}_{An} \vert J/\psi \rangle (1 + \mathcal {O}(v^2)),
\end{split}
\end{align}
which can generate relation between SGF matrix element and NRQCD matrix elements with $\mathcal {O}(v^2)$ denoting contributions from operators with explicit gauge fields. Actually, we can generate even more relations by applying the two factorization formulas to any well defined QCD quantity $W$,
\begin{align}
\begin{split}
W= \hat{W}\overline{R}^{\CScSa *}_{J/\psi} (1 + \mathcal {O}(v^2))
=&  \sum_{n} w_n(\CScSa) \langle 0 \vert \mathcal{O}_{An} \vert J/\psi \rangle (1 + \mathcal {O}(v^2)).
\end{split}
\end{align}
For example, if we choose $W=\frac{1}{4}(M+2m)(M-2m)\overline{R}^{\CScSa *}_{J/\psi}$, we have $\hat{W}=\frac{1}{4}(M+2m)(M-2m)$. To determine corresponding $w_n$, we replace $J/\psi$ by a $c\bar c$ pair with invariant mass $M$~\footnote{Note that $M$ is exactly the mass of $J/\psi$. This is very important, or else $M$ is $W$ needs to be expanded.}. Using the nonrelativistic
expansion formulas given in~\cite{Braaten:1996rp}, we have
\begin{align}
\begin{split}
 \overline{R}^{\CScSa*,(0)}_{c \bar c} =& \frac{1}{\sqrt{N_c}}\frac{\sqrt{M}}{M+2m} \bar v(p_{\bar c})\frac{M-\slashed{P}}{2M}{\epsilon}_{S_z}^{*\mu} \gamma_\mu\frac{M+\slashed{P}}{2M} u(p_{ c})\\
 =& -\frac{1}{2\sqrt{M}\sqrt{N_c}}  \eta^\dag \boldsymbol{\sigma} \cdot \boldsymbol{\epsilon}_{S_z}^{*} \xi,\\
\langle 0 \vert \mathcal{O}_{An} \vert c \bar c \rangle^{(0)}= & \boldsymbol{q}^{2n} \eta^\dag \boldsymbol{\sigma} \cdot \boldsymbol{\epsilon}_{S_z}^{*} \xi.
\end{split}
\end{align}
Here we used nonrelativistic normalization for the spinors $u$ and $v$.
Then we obtain
\begin{align}
\begin{split}
w_n^{(0)}(\CScSa)=& \bigg[ \frac{1}{n!} \bigg(\frac{\partial}{\partial \boldsymbol{q}^2} \bigg)^{n}\bigg( -\frac{\sqrt{2M}}{2\sqrt{M}\sqrt{N_c}} \boldsymbol{q}^{2} \bigg) \bigg]
\bigg\vert_{\boldsymbol{q}^2=0} \\
=&-\frac{1}{\sqrt{2N_c}} \delta_{n1} .
\end{split}
\end{align}
Where the extra factor $\sqrt{2M}$ in the first line appears because NRQCD matrix elements have nonrelativistic normalization, while SGF matrix elements have relativistic normalization.
As both SGF matrix elments and NRQCD matrix elements keep the same low energy physics and renormalized in the same way, the coefficients $w_n(\CScSa)$ should vanish at higher orders in $\alpha_s$, i.e.
\begin{align}
\begin{split}
w_n^{(i)}(\CScSa)=0,  \quad\quad i\geq 1.
\end{split}
\end{align}
Thus we get a relation
\begin{align}
\begin{split}
 \frac{1}{4}(M+2m)(M-2m)\overline{R}^{\CScSa *}_{J/\psi}
=&   -\frac{1}{\sqrt{2N_c}} \langle 0 \vert \mathcal{O}_{A1} \vert J/\psi \rangle (1 + \mathcal {O}(v^2)).
\end{split}
\end{align}

Similarly, by choosing $W=[\frac{1}{4}(M+2m)(M-2m)]^n\overline{R}^{\CScSa *}_{J/\psi}$, we can obtain
\begin{align}
\begin{split}\label{eq:Eq:ME-relation}
[\frac{1}{4}(M+2m)(M-2m)]^n\overline{R}^{\CScSa *}_{J/\psi}
=&   -\frac{1}{\sqrt{2N_c}} \langle 0 \vert \mathcal{O}_{An} \vert J/\psi \rangle (1 + \mathcal {O}(v^2)).
\end{split}
\end{align}
which provides complete relations between the SGF matrix element and NRQCD matrix elements. Using these relations, we obtain
\begin{align}
\begin{split}\label{eq:Eq:ME-relationn}
\frac{\langle 0 \vert \mathcal{O}_{An} \vert J/\psi \rangle}{\langle 0 \vert \mathcal{O}_{A0} \vert J/\psi \rangle}=&   [\frac{1}{4}(M+2m)(M-2m)]^n (1 + \mathcal {O}(v^2)),
\end{split}
\end{align}
which agrees with Eq.~\eqref{eq:general-Gremm-Kapustin}.
We thus have proved the generalized Gremm-Kapustin relations by using the equivalence between SGF and NRQCD factorization. Based on our proof, it is clear that the $\mathcal {O}(v^2)$ terms in the relations are contributions from operators with explicit gauge fields.

\section{Deducing SGF from NRQCD factorization}\label{sec:general}

\subsection{Exclusive processes}

Because the equivalence between SGF and NRQCD factorization gives rise to the generalized Gremm-Kapustin relations, it implies that Gremm-Kapustin-like relations are the key to relate NRQCD to SGF. Indeed, due to these relation, the introduction of nonperturbative quantities $\langle 0 \vert\mathcal{O}_{An} \vert J/\psi \rangle$ with $ n\geq 1$ is unnecessary for exclusive processes. The dominant contributions of these quantities are purely kinematic that can be taken into account by coefficients of $\langle 0 \vert\mathcal{O}_{A0} \vert J/\psi \rangle$. Thus the NRQCD factorization formula can be resummed to obtain
\begin{align}
\begin{split}\label{eq:NR-SGF-1}
\mathcal{A}^{J/\psi \to e^+e^- }
=&   \biggr( \hat{\mathcal{A}}^{\CScSa,(0)} + \hat{\mathcal{A}}^{\CScSa,(1)} +\cdots\biggr) \frac{-1}{\sqrt{2N_c}}  \langle 0 \vert\mathcal{O}_{A0} \vert J/\psi \rangle (1 + \mathcal {O}(v^2)),
\end{split}
\end{align}
which is exactly the SGF formula, noticing the Eq.~\eqref{eq:Eq:ME-relation} which relates $\langle 0 \vert\mathcal{O}_{A0} \vert J/\psi \rangle$  to $\overline{R}^{\CScSa *}_{J/\psi}$.

Now let us show generally that SGF can be deduced from NRQCD at the operator level. The equations of motion of heavy quark fields in NRQCD are given by~\cite{Caswell:1985ui}
\begin{align}
	\begin{split}
\left( i D_0 -\frac{{\bm D}^2}{2m}+\cdots \right)\psi&=0,
\end{split}
\end{align}
with similar equation for $\chi$ field.
Because we are not interested in gluon fields, we can replace $D_0$ by $\nabla_0$ and ${\bm D}$ by ${\bm \nabla}$. The equations of motion are usually used to replace operators involving $\nabla_0$ by operators involving ${\bm \nabla}^2$ in NRQCD. Then only spacial components ${\bm \nabla}$ appear in NRQCD operators, which can be decomposed to relative derivative and total derivative when it acts on quark-antiquark bilinear operators. For example, beginning from the bilinear operator $\chi^\dagger\psi$, in NRQCD one can construct more operators by adding relative derivative to obtain $\chi^\dagger \overleftrightarrow{\bm \nabla}^2 \psi$, or total derivative to obtain ${\bm \nabla}^2(\chi^\dagger \psi)$, or their combinations.

However, the equations of motion can also be used to replace relative derivatives $\overleftrightarrow{\bm \nabla}^2$ and $\overleftrightarrow{ \nabla}_0$ by total derivatives~\footnote{During the matching procedure in SGF, We indeed using onshell conditions to fix relative momentum. See discussion after Eq.~\eqref{eq:onshel}.}, which results in the following matrix elements for exclusive processes
\begin{align}
\begin{split}
\langle0| & \nabla_0^{n_1} {\bm \nabla}^{2n_2}(\chi^\dagger \psi)| {\cal Q}\rangle\,,
\end{split}
\end{align}
with non-negative integers for $n_1$ and $n_2$. As we are working in the rest frame of $\cal Q$, by using integration by parts one finds that $\nabla_0$ gives rise to quarkonium mass $M$ and ${\bm \nabla}$ vanishes. Therefore, in a factorization formula the matrix element with $n_1=n_2=0$ is enough to take care of all contributions in this series of matrix elements, although short-distance hard parts will depend on heavy quark mass $m$ as while as quarkonium mass $M$. This is nothing but the SGF formula. It is also clear that the SGF resums a series of power corrections originated from kinematic effects in NRQCD.

\subsection{Inclusive processes}

For inclusive quarkonium processes, we can also use equations of motion to decompose NRQCD matrix elements by
\begin{align}
\begin{split}
\langle {\cal Q}+X| & \nabla_0^{n_1} {\bm \nabla}^{2n_2}(\psi^\dagger  \chi)|0\rangle\,.
\end{split}
\end{align}
Using integration by parts we can eliminate all matrix elements except $n_1=n_2=0$, but with short-distance hard parts depending on $P^2$, $P\cdot P_X$ and $P_X^2$, where $P_X$ is the momentum of unobserved particles $X$. The final result is the SGF formula for inclusive quarkonium processes. It again resums a series of power corrections originated from kinematic effects in NRQCD.

Because short-distance hard parts depend on $P_X$, nonperturbative matrix elements, soft gluon distributions, must be also functions of $P_X$~\cite{Ma:2017xno}. Note the difference between soft gluon distributions and shape functions introduced at endpoint region~\cite{Beneke:1997qw,Fleming:2003gt,Fleming:2006cd,Leibovich:2007vr}. The purpose of shape functions is to resum large logarithms at endpoint region  in NRQCD factorization framework, which are defined at fixed power in relativistic expansion (usually leading power). The SGF with soft gluon distributions are aimed at resumming a power series of relativistic expansion, which can both be applied at endpoint region and regions away from that. If SGF is applied at endpoint region, large logarithms can be naturally resummed by renormalization group equations of soft gluon distribution. Detailed discussion of this topic will be presented in a separate work~\cite{chenma2020}.

\section{Summary}\label{sec:Conclusion}
In summary, taking $\Gamma(J/\psi \to e^+e^-)$ as an example, we demonstrated that the SGF is equivalent to the NRQCD factorization for heavy quarkonium production or decay. 
We also shown that the SGF can be deduced from NRQCD effective field theory at the operator level by using equations of motion. To achieve the above conclusion, we introduced a new regulator and defined SGF matrix elements rigorously. 
Based on the equivalence between the two factorizations, we derived explicit relations between SGF matrix elements and NRQCD matrix elements and proved the generalized Gremm-Kapustin relations. 

The results obtained in this paper means that, for any given process, the SGF and the NRQCD factorization are either both valid or both violated. Therefore, the SGF is valid to all orders in perturbation theories for many processes where NRQCD factorization have been proved . This provides a solid theoretical foundation for the SGF.
Comparing with the NRQCD factorization, the SGF effectively resums a subset of relativistic correction terms originated from kinematic effects, which can reduce theoretical
uncertainties and thus may provide a better description of experimental data.

\section*{Acknowledgments}
\label{sec:acknowledgments}

We would like to thank K.T. Chao and C. Meng for useful discussions. The work is supported by the National Natural Science Foundation of China (Grants No. 11875071, No. 11975029) and the China Postdoctoral Science Foundation under Grant No.2018M631234.


\appendix

\section{Full QCD results calculated by regions}\label{sec:app}

In this appendix, we use the method of regions to calculate the amplitude $\mathcal{A}^{c\bar{c}(\CScSa) \to e^+e^-}$ at one-loop order.
According to Eqs.~\eqref{eq:split} and~\eqref{eq:exp}, for original integral $F$ one has
\begin{align}
\begin{split}
F
=& \int_{k\in D_h} d^dk \bigg\{I - \bigg[ T^{(h,s)} + T^{(h,p)}- T^{(h,s,p)} \bigg]I \bigg\} + \int d^dk \bigg\{T^{(s)}+ T^{(p)} - T^{(s,p)}\bigg\}I \\
=& \int d^dk \bigg\{T^{(h)}-T^{(h,s)} - T^{(h,p)}+ T^{(h,s,p)} \bigg\}I + \int d^dk \bigg\{T^{(s)}+ T^{(p)} - T^{(s,p)}\bigg\}I.
\end{split}
\end{align}
Defining
\begin{align}
\begin{split}
F^{(i,j,\cdots)} \equiv \int d^dk T^{(i,j,\cdots)} I,
\end{split}
\end{align}
 we have
\begin{align}
\begin{split} \label{eq:F-formula}
F
=& \bigg\{F^{(h)}-F^{(h,s)} - F^{(h,p)}+ F^{(h,s,p)} \bigg\}+ \bigg\{F^{(s)}+ F^{(p)} - F^{(s,p)}\bigg\}.
\end{split}
\end{align}
As the first term on the r.h.s can be defined by integrals in hard domain, it is infrared safe.

Now we apply the formula Eq.~\eqref{eq:F-formula} to calculate the QCD correction to $\mathcal {A}^\mu_{c+\bar c}$.
We first consider the vertex correction. As showed in Ref.~\cite{Bodwin:2008vp}, the vertex correction can be expressed in terms
of elementary integrals $I_{111}, I_{011}, I_{-111}, I_{010}$ and $I_{110}$, with $I_{abc}$
defined as
\begin{align}
\begin{split}
I_{abc} \equiv \mu^{2\epsilon} \int\frac{d^dk}{(2\pi)^d} \frac{1}{[k^2+i0^+]^a[k^2+2p_{ c}\cdot k +i0^+]^b[k^2-2p_{\bar c}\cdot k +i0^+]^c}.
\end{split}
\end{align}
According to Eq.~\eqref{eq:F-formula}, $I_{abc}$ can be expressed as
\begin{align}
\begin{split}\label{eq:Iabc}
I_{abc}
=& \left\{ I_{abc}^{(h)}-I_{abc}^{(h,s)}-I_{abc}^{(h,p)}  + I_{abc}^{(h,s,p)} \right\} +  \left\{I_{abc}^{(s)}+ I_{abc}^{(p)}
-I_{abc}^{(s,p)} \right\}  .
\end{split}
\end{align}
The overlap contributions $I_{abc}^{(h,s)}, I_{abc}^{(h,p)}, I_{abc}^{(h,s,p)}, I_{abc}^{(s,p)}$ and soft region contribution $I_{abc}^{(s)}$ are scaleless integrals and can be set to zero if there is no infrared divergence.
The calculation of $I_{abc}^{(s)}-I_{abc}^{(s,p)}$ and $I_{abc}^{(h,s)}-I_{abc}^{(h,s,p)}$ is similar to $\overline{R}^{\CScSa*,(s)}_{c\bar{c}(\CScSa)}
-\overline{R}^{\CScSa*,(s,p)}_{c\bar{c}(\CScSa)}$, one has
\begin{subequations}\label{eq:I-s}
\begin{align}
I_{111}^{(s)}-I_{111}^{(s,p)}=& I_{111}^{(h,s)}-I_{111}^{(h,s,p)}= \frac{i}{(4\pi)^2}\frac{1}{M^2} \left[ 2L(\delta) \left(\frac{1}{\epsilon_{UV}}-\frac{1}{\epsilon_{IR}} \right)   \right ],\\
I_{011}^{(s)}-I_{011}^{(s,p)}=& I_{011}^{(h,s)}-I_{011}^{(h,s,p)}=0,\\
I_{-11}^{(s)}-I_{-11}^{(s,p)}=& I_{-11}^{(h,s)}-I_{-11}^{(h,s,p)}=0,\\
I_{010}^{(s)}-I_{010}^{(s,p)}=& I_{010}^{(h,s)}-I_{010}^{(h,s,p)}=0,\\
I_{110}^{(s)}-I_{110}^{(s,p)}=& I_{110}^{(h,s)}-I_{110}^{(h,s,p)}=0.
\end{align}
\end{subequations}
Then we consider the contributions $I_{abc}^{(h,p)}$. Let's take $I_{111}^{(h,p)}$ for example. The expanded integral reads
\begin{align}
\begin{split}
I_{111}^{(h,p)}=&
 \mu^{2\epsilon} \sum_{j_1,\cdots, j_5=0}   \frac{j_{12}!j_{34}!}{j_1!\cdots j_4! }   \int\frac{d^dk}{(2\pi)^d} \frac{(-k_0^2)^{j_{135}}(2 \boldsymbol{q} \cdot \boldsymbol{k})^{j_{24}}}{[- \boldsymbol{k}^2]^{1+j_5}[- \boldsymbol{k}^2 + k_0p_0]^{1+j_{12}}[- \boldsymbol{k}^2 - k_0p_0]^{1+j_{34}}}.
\end{split}
\end{align}
These integrals are scaleless ans infrared safe and thus vanish. Similarly we have
\begin{align}\label{eq:I-hp}
I_{111}^{(h,p)}
=
I_{011}^{(h,p)}
=
I_{-111}^{(h,p)}
=
I_{010}^{(h,p)}
=
I_{110}^{(h,p)}
=0.
\end{align}
The remaining contributions $I_{abc}^{(h)}$ and $I_{abc}^{(p)}$ are given by
\begin{subequations}\label{eq:I-h-p}
\begin{align}
I_{111}^{(h)}
=& \frac{i}{(4\pi)^2}\frac{1}{M^2} \left[ -2L(\delta)\left(\frac{1}{\epsilon} + \log \frac{4\pi \mu^2 e^{-\gamma_E}}{ m^2 } \right) + 4K(\delta)  \right ],\\
I_{011}^{(h)}
=& \frac{i}{(4\pi)^2}\left[\frac{1}{\epsilon} + \log \frac{4\pi \mu^2 e^{-\gamma_E}}{ m^2 } +2- 2\delta^2L(\delta) \right ],\\
I_{-111}^{(h)}
=& \frac{i}{(4\pi)^2}\frac{M^2}{4} \left[ (1-3\delta^2)\left(\frac{1}{\epsilon} + \log \frac{4\pi \mu^2 e^{-\gamma_E}}{ m^2 } +1 \right)-2\delta^2 +4\delta^4L(\delta) \right ],\\
I_{010}^{(h)}
=& \frac{i}{(4\pi)^2} m^2 \left[\frac{1}{\epsilon} + \log \frac{4\pi \mu^2 e^{-\gamma_E}}{ m^2 } +1\right ],\\
I_{110}^{(h)}
=& \frac{i}{(4\pi)^2}  \left[\frac{1}{\epsilon} + \log \frac{4\pi \mu^2 e^{-\gamma_E}}{ m^2 } +2\right ],\\
I_{111}^{(p)}
=& \frac{i}{(4\pi)^2}\frac{1}{M^2} \left( -\frac{\pi^2}{\delta}  \right ),\\
I_{011}^{(p)}
=&
I_{-111}^{(p)}
=
I_{010}^{(p)}
=
I_{110}^{(p)}
=0.
\end{align}
\end{subequations}
It should be noted that ultraviolet and infrared divergences in $I_{abc}^{(h)}$ are not well distinguished. But when overlap contributions $I_{abc}^{(h,s)}, I_{abc}^{(h,p)}, I_{abc}^{(h,p,s)}$ are subtracted from $I_{abc}^{(h)}$, infrared divergent parts of in the hard region can be removed. Then the $1/\epsilon$
divergences in $I_{abc}^{(h)}-I_{abc}^{(h,s)}-I_{abc}^{(h,p)}  + I_{abc}^{(h,s,p)}$ are converted to $1/\epsilon_{UV}$ divergences. Inserting Eqs.~\eqref{eq:I-s}, \eqref{eq:I-hp} and \eqref{eq:I-h-p} into \eqref{eq:Iabc}, we find that results for elementary integrals are consistent with those in Refs.~\cite{Bodwin:2008vp, Lee:2007kg}.

We also need to calculate heavy-quark wave-function renormalization $Z_Q$, which given by
\begin{align}
\begin{split}\label{wave-function-Z}
Z_Q=1+\frac{p_c^\mu}{m}\frac{\partial \Sigma(p_c)}{\partial p_c^\mu} \biggr\vert_{\slashed{p}_{{c}}=m} + O(\alpha_s^2),
\end{split}
\end{align}
where
\begin{align}
\begin{split}\label{wave-function}
\frac{p_c^\mu}{m}\frac{\partial \Sigma(p_c)}{\partial p_c^\mu} \biggr\vert_{\slashed{p}_{{c}}=m} =& -ig_s^2 C_F  \biggr[ -dT_{11}+2(T_{02}-2m^2T_{12})- \frac{(2-d)\slashed{P}}{2m}(T_{11}^0-T_{02}^0 +2m^2T_{12}^0 )
\\
&-\frac{(2-d)\slashed{q}}{m}(T_{11}^1-T_{02}^1 +2m^2T_{12}^1 )  \biggr].
\end{split}
\end{align}
Here we have introduced integrals $T_{ab}, T_{ab}^0, T_{ab}^1$, which are defined by
\begin{subequations}
\begin{align}
T_{ab} \equiv& \mu^{2\epsilon} \int\frac{d^dk}{(2\pi)^d} \frac{1}{[k^2+i0^+]^a[k^2+2p_{ c}\cdot k +i0^+]^b},\\
T_{ab}^0 \equiv& \mu^{2\epsilon} \int\frac{d^dk}{(2\pi)^d} \frac{2}{P^2}\frac{P\cdot k}{[k^2+i0^+]^a[k^2+2p_{ c}\cdot k +i0^+]^b} ,\\
T_{ab}^1 \equiv& \mu^{2\epsilon} \int\frac{d^dk}{(2\pi)^d} \frac{1}{q^2}\frac{q\cdot k}{[k^2+i0^+]^a[k^2+2p_{ c}\cdot k +i0^+]^b}.
\end{align}
\end{subequations}
Similarly, we can derive
\begin{subequations}\label{T-h}
\begin{align}
T_{11}^{(h)} =& I_{110}^{(h)}= \frac{i}{(4\pi)^2}  \left[\frac{1}{\epsilon} + \log \frac{4\pi \mu^2 e^{-\gamma_E}}{ m^2 } +2\right ],\\
T_{02}^{(h)} =& \frac{i}{(4\pi)^2}  \left[\frac{1}{\epsilon} + \log \frac{4\pi \mu^2 e^{-\gamma_E}}{ m^2 } \right ],\\
T_{12}^{(h)} =& \frac{i}{(4\pi)^2} \frac{1}{2m^2} \left[\frac{1}{\epsilon} + \log \frac{4\pi \mu^2 e^{-\gamma_E}}{ m^2 } \right ],\\
T_{11}^{0,(h)} =& T_{11}^{1,(h)}= - \frac{1}{2}\frac{i}{(4\pi)^2}  \left[\frac{1}{\epsilon} + \log \frac{4\pi \mu^2 e^{-\gamma_E}}{ m^2 } +1\right ],\\
T_{02}^{0,(h)} =& T_{02}^{1,(h)}= - \frac{i}{(4\pi)^2}  \left[\frac{1}{\epsilon} + \log \frac{4\pi \mu^2 e^{-\gamma_E}}{ m^2 } \right ],\\
T_{12}^{0,(h)} =& T_{12}^{1,(h)}=  \frac{i}{(4\pi)^2} \frac{1}{m^2}.
\end{align}
\end{subequations}
For the remaining contributions, only $T_{12}^{(s)}-T_{12}^{(s,p)}$ and $T_{12}^{(h,s)}-T_{12}^{(h,s,p)}$ are non-vanish, which read
\begin{align}
\begin{split}\label{T-s}
T_{12}^{(s)}-T_{12}^{(s,p)} =& T_{12}^{(h,s)}-T_{12}^{(h,s,p)}= - \frac{i}{(4\pi)^2} \frac{1}{2m^2} \left(\frac{1}{\epsilon_{UV}} - \frac{1}{\epsilon_{IR}}  \right ).
\end{split}
\end{align}
Making use of Eqs.~\eqref{wave-function-Z}, \eqref{wave-function}, \eqref{T-h} and \eqref{T-s}, we obtain
\begin{align}
\begin{split}
Z_Q=1+\frac{\alpha_sC_F}{4\pi} \left[ -\frac{1}{\epsilon_{UV}} - \frac{2}{\epsilon_{IR}} - 3\log \frac{4\pi \mu^2 e^{-\gamma_E}}{ m^2 } -4 \right].
\end{split}
\end{align}
This is agree with the result in Ref.~\cite{Braaten:1995ej}.

Based on the above results, one can reproduce the full QCD result in Eq.~\eqref{eq:dec-coefficient} by summing the contributions from different domains:
\begin{subequations}\label{eq:G-H}
\begin{align}
\mathcal {G}^{(1)}
=& \left\{\mathcal {G}^{(1),(h)} -\mathcal {G}^{(1),(h,s)} -\mathcal {G}^{(1),(h,p)} +\mathcal {G}^{(1),(h,s,p)} \right\} + \left\{\mathcal {G}^{(1),(s)} +\mathcal {G}^{(1),(p)} - \mathcal {G}^{(1),(s,p)} \right\}, \\
\mathcal {H}^{(1)}
=& \left\{ \mathcal {H}^{(1),(h)} -\mathcal {H}^{(1),(h,s)} -\mathcal {H}^{(1),(h,p)} +\mathcal {H}^{(1),(h,s,p)} \right\} + \left\{ \mathcal {H}^{(1),(s)} +\mathcal {H}^{(1),(p)} - \mathcal {H}^{(1),(s,p)} \right\},
\end{align}
\end{subequations}
with
\begin{subequations}\label{eq:hp-region}
\begin{align}
\mathcal {G}^{(1),(h)}=& \frac{\alpha_s C_F}{4\pi} \biggr[ (2 (1+\delta^2)L(\delta) -2 ) \biggr(\frac{1}{\epsilon} + \log \frac{4\pi \mu^2 e^{-\gamma_E}}{ m^2 } \biggr) + 6\delta^2L(\delta) -4(1+\delta^2)K(\delta)-4
\biggr],\\
\mathcal {G}^{(1),(p)}
=& \frac{\alpha_s C_F}{4\pi}  (1+\delta^2 )\frac{\pi^2}{\delta},\\
\mathcal {G}^{(1),(h,p)}
=& 0,\\
\mathcal {G}^{(1),(s)}-\mathcal {G}^{(1),(s,p)} =&\mathcal {G}^{(1),(h,s)}-\mathcal {G}^{(1),(h,s,p)}
= - \frac{\alpha_s C_F}{4\pi}  (2(1+\delta^2 )L(\delta) -2) \biggr(\frac{1}{\epsilon_{UV}}-\frac{1}{\epsilon_{IR}}\biggr), \\
\mathcal {H}^{(1),(h)}
=& \frac{\alpha_s C_F}{4\pi}  \frac{1-\delta^2 }{m}2L(\delta),\\
\mathcal {H}^{(1),(p)}
=& 0, \\
\mathcal {H}^{(1),(h,p)}
=& 0, \\
\mathcal {H}^{(1),(s)}-\mathcal {H}^{(1),(s,p)} =&\mathcal {H}^{(1),(h,s)}-\mathcal {H}^{(1),(h,s,p)}
= 0.
\end{align}
\end{subequations}
Substituting Eq.~\eqref{eq:G-H} into Eq.~\eqref{eq:amplitude-result},  $\mathcal{A}^{c\bar{c}(\CScSa)\to e^+e^-,(1)}$ can be expressed as
\begin{align}
\begin{split}\label{eq:amplitude-re}
\mathcal{A}^{c\bar{c}(\CScSa)\to e^+e^-,(1)}
=& \bigg\{\mathcal {A}^{(h)}
-\mathcal {A}^{(h,s)}-\mathcal {A}^{(h,p)}
+\mathcal {A}^{(h,s,p)} \bigg\}   + \bigg\{ \mathcal {A}^{(s)}+ \mathcal {A}^{(p)}-\mathcal {A}^{(s,p)} \bigg\},
\end{split}
\end{align}
with $\mathcal {A}^{(i,j,\cdots)}$ determined by $\mathcal {G}^{(1),(i,j,\cdots)}$
and $\mathcal {H}^{(1),(i,j,\cdots)}$.


\begin{thebibliography}{10}

\bibitem{Bodwin:1994jh}
G.~T. Bodwin, E.~Braaten, and G.~P. Lepage, {\it {Rigorous QCD analysis of
  inclusive annihilation and production of heavy quarkonium}},
  \href{http://dx.doi.org/10.1103/PhysRevD.55.5853,
  10.1103/PhysRevD.51.1125}{{\em Phys. Rev.} {\bfseries D51} (1995) 1125--1171}
  [\href{http://arxiv.org/abs/hep-ph/9407339}{{\ttfamily hep-ph/9407339}}]
  [\href{http://inspirehep.net/search?p=find+Bodwin:1994jh}{{\ttfamily
  InSPIRE}}].
[Erratum: Phys. Rev.D55,5853(1997)].

\bibitem{Caswell:1985ui}
W.~Caswell and G.~Lepage, {\it {Effective Lagrangians for Bound State Problems
  in QED, QCD, and Other Field Theories}},
\href{http://dx.doi.org/10.1016/0370-2693(86)91297-9}{{\em Phys.Lett.}
  {\bfseries B167} (1986) 437}
  [\href{http://inspirehep.net/search?p=find+Caswell:1985ui}{{\ttfamily
  InSPIRE}}].

\bibitem{Nayak:2005rt}
G.~C. Nayak, J.-W. Qiu, and G.~Sterman, {\it {Fragmentation, NRQCD and NNLO
  factorization analysis in heavy quarkonium production}},
\href{http://dx.doi.org/10.1103/PhysRevD.72.114012}{{\em Phys.Rev.} {\bfseries
  D72} (2005) 114012} [\href{http://arxiv.org/abs/hep-ph/0509021}{{\ttfamily
  hep-ph/0509021}}]
  [\href{http://inspirehep.net/search?p=find+Nayak:2005rt}{{\ttfamily
  InSPIRE}}].

\bibitem{Nayak:2005rw}
G.~C. Nayak, J.-W. Qiu, and G.~Sterman, {\it {Fragmentation, factorization and
  infrared poles in heavy quarkonium production}},
\href{http://dx.doi.org/10.1016/j.physletb.2005.03.031}{{\em Phys.Lett.}
  {\bfseries B613} (2005) 45--51}
  [\href{http://arxiv.org/abs/hep-ph/0501235}{{\ttfamily hep-ph/0501235}}]
  [\href{http://inspirehep.net/search?p=find+Nayak:2005rw}{{\ttfamily
  InSPIRE}}].

\bibitem{Nayak:2006fm}
G.~C. Nayak, J.-W. Qiu, and G.~Sterman, {\it {NRQCD Factorization and
  Velocity-dependence of NNLO Poles in Heavy Quarkonium Production}},
\href{http://dx.doi.org/10.1103/PhysRevD.74.074007}{{\em Phys.Rev.} {\bfseries
  D74} (2006) 074007} [\href{http://arxiv.org/abs/hep-ph/0608066}{{\ttfamily
  hep-ph/0608066}}]
  [\href{http://inspirehep.net/search?p=find+Nayak:2006fm}{{\ttfamily
  InSPIRE}}].

\bibitem{Mangano:1996kg}
M.~L. Mangano and A.~Petrelli, {\it {NLO quarkonium production in hadronic
  collisions}},
\href{http://dx.doi.org/10.1142/S0217751X97002048}{{\em Int. J. Mod. Phys.}
  {\bfseries A12} (1997) 3887--3897}
  [\href{http://arxiv.org/abs/hep-ph/9610364}{{\ttfamily hep-ph/9610364}}]
  [\href{http://inspirehep.net/search?p=find+Mangano:1996kg}{{\ttfamily
  InSPIRE}}].

\bibitem{Ma:2017xno}
Y.-Q. Ma and K.-T. Chao, {\it {New factorization theory for heavy quarkonium
  production and decay}},
\href{http://dx.doi.org/10.1103/PhysRevD.100.094007}{{\em Phys. Rev.}
  {\bfseries D100} (2019) 094007}
  [\href{http://arxiv.org/abs/1703.08402}{{\ttfamily arXiv:1703.08402}}]
  [\href{http://inspirehep.net/search?p=find+Ma:2017xno}{{\ttfamily InSPIRE}}].

\bibitem{Li:2019ncs}
R.~Li, Y.~Feng, and Y.-Q. Ma, {\it {Exclusive quarkonium production or decay in
  soft gluon factorization}},
  [\href{http://arxiv.org/abs/1911.05886}{{\ttfamily arXiv:1911.05886}}]
  [\href{http://inspirehep.net/search?p=find+Li:2019ncs}{{\ttfamily InSPIRE}}].

\bibitem{Bodwin:2008vp}
G.~T. Bodwin, H.~S. Chung, J.~Lee, and C.~Yu, {\it {Order-$\alpha_s$
  corrections to the quarkonium electromagnetic current at all orders in the
  heavy-quark velocity}},
\href{http://dx.doi.org/10.1103/PhysRevD.79.014007}{{\em Phys.Rev.} {\bfseries
  D79} (2009) 014007} [\href{http://arxiv.org/abs/0807.2634}{{\ttfamily
  arXiv:0807.2634}}]
  [\href{http://inspirehep.net/search?p=find+Bodwin:2008vp}{{\ttfamily
  InSPIRE}}].

\bibitem{Beneke:1997zp}
M.~Beneke and V.~A. Smirnov, {\it {Asymptotic expansion of Feynman integrals
  near threshold}},
\href{http://dx.doi.org/10.1016/S0550-3213(98)00138-2}{{\em Nucl.Phys.}
  {\bfseries B522} (1998) 321--344}
  [\href{http://arxiv.org/abs/hep-ph/9711391}{{\ttfamily hep-ph/9711391}}]
  [\href{http://inspirehep.net/search?p=find+Beneke:1997zp}{{\ttfamily
  InSPIRE}}].

\bibitem{Jantzen:2011nz}
B.~Jantzen, {\it {Foundation and generalization of the expansion by regions}},
  \href{http://dx.doi.org/10.1007/JHEP12(2011)076}{{\em JHEP} {\bfseries 12}
  (2011) 076} [\href{http://arxiv.org/abs/1111.2589}{{\ttfamily
  arXiv:1111.2589}}]
  [\href{http://inspirehep.net/search?p=find+Jantzen:2011nz}{{\ttfamily
  InSPIRE}}].

\bibitem{Liu:2019iml}
H.-Y. Liu, Y.-Q. Ma, and K.-T. Chao, {\it {Improvement for Color Glass
  Condensate factorization: single hadron production in pA collisions at
  next-to-leading order}},
  \href{http://dx.doi.org/10.1103/PhysRevD.100.071503}{{\em Phys. Rev. D}
  {\bfseries 100} (2019) 071503}
  [\href{http://arxiv.org/abs/1909.02370}{{\ttfamily arXiv:1909.02370}}]
  [\href{http://inspirehep.net/search?p=find+Liu:2019iml}{{\ttfamily
  InSPIRE}}].

\bibitem{Bodwin:2007fz}
G.~T. Bodwin, H.~S. Chung, D.~Kang, J.~Lee, and C.~Yu, {\it {Improved
  determination of color-singlet nonrelativistic QCD matrix elements for S-wave
  charmonium}},
\href{http://dx.doi.org/10.1103/PhysRevD.77.094017}{{\em Phys. Rev.} {\bfseries
  D77} (2008) 094017} [\href{http://arxiv.org/abs/0710.0994}{{\ttfamily
  arXiv:0710.0994}}]
  [\href{http://inspirehep.net/search?p=find+Bodwin:2007fz}{{\ttfamily
  InSPIRE}}].

\bibitem{Lee:2018aoz}
J.~Lee, J.~H. Ee, U.-R. Kim, and C.~Yu, {\it {Quarkonium physics: NRQCD
  factorization formula for $J/\psi \to e^+ e^-$}},
\href{http://dx.doi.org/10.23730/CYRSP-2018-002.69}{{\em CERN Yellow Rep.
  School Proc.} {\bfseries 2} (2018) 69--85}
  [\href{http://inspirehep.net/search?p=find+Lee:2018aoz}{{\ttfamily
  InSPIRE}}].

\bibitem{Bodwin:2006dn}
G.~T. Bodwin, D.~Kang, and J.~Lee, {\it {Potential-model calculation of an
  order-$v^2$ NRQCD matrix element}},
\href{http://dx.doi.org/10.1103/PhysRevD.74.014014}{{\em Phys.Rev.} {\bfseries
  D74} (2006) 014014} [\href{http://arxiv.org/abs/hep-ph/0603186}{{\ttfamily
  hep-ph/0603186}}]
  [\href{http://inspirehep.net/search?p=find+Bodwin:2006dn}{{\ttfamily
  InSPIRE}}].

\bibitem{Gremm:1997dq}
M.~Gremm and A.~Kapustin, {\it {Annihilation of S wave quarkonia and the
  measurement of alpha-s}},
\href{http://dx.doi.org/10.1016/S0370-2693(97)00744-2}{{\em Phys. Lett.}
  {\bfseries B407} (1997) 323--330}
  [\href{http://arxiv.org/abs/hep-ph/9701353}{{\ttfamily hep-ph/9701353}}]
  [\href{http://inspirehep.net/search?p=find+Gremm:1997dq}{{\ttfamily
  InSPIRE}}].

\bibitem{Braaten:1996rp}
E.~Braaten and Y.-Q. Chen, {\it {Dimensional regularization in quarkonium
  calculations}},
\href{http://dx.doi.org/10.1103/PhysRevD.55.2693}{{\em Phys.Rev.} {\bfseries
  D55} (1997) 2693--2707}
  [\href{http://arxiv.org/abs/hep-ph/9610401}{{\ttfamily hep-ph/9610401}}]
  [\href{http://inspirehep.net/search?p=find+Braaten:1996rp}{{\ttfamily
  InSPIRE}}].

\bibitem{Beneke:1997qw}
M.~Beneke, I.~Z. Rothstein, and M.~B. Wise, {\it {Kinematic enhancement of
  nonperturbative corrections to quarkonium production}},
\href{http://dx.doi.org/10.1016/S0370-2693(97)00832-0}{{\em Phys. Lett.}
  {\bfseries B408} (1997) 373--380}
  [\href{http://arxiv.org/abs/hep-ph/9705286}{{\ttfamily hep-ph/9705286}}]
  [\href{http://inspirehep.net/search?p=find+Beneke:1997qw}{{\ttfamily
  InSPIRE}}].

\bibitem{Fleming:2003gt}
S.~Fleming, A.~K. Leibovich, and T.~Mehen, {\it {Resumming the color octet
  contribution to $e^{+} e^{-} \to J/\psi$ + $X$}},
\href{http://dx.doi.org/10.1103/PhysRevD.68.094011}{{\em Phys.Rev.} {\bfseries
  D68} (2003) 094011} [\href{http://arxiv.org/abs/hep-ph/0306139}{{\ttfamily
  hep-ph/0306139}}]
  [\href{http://inspirehep.net/search?p=find+Fleming:2003gt}{{\ttfamily
  InSPIRE}}].

\bibitem{Fleming:2006cd}
S.~Fleming, A.~K. Leibovich, and T.~Mehen, {\it {Resummation of Large Endpoint
  Corrections to Color-Octet $J/\psi$ Photoproduction}},
\href{http://dx.doi.org/10.1103/PhysRevD.74.114004}{{\em Phys. Rev.} {\bfseries
  D74} (2006) 114004} [\href{http://arxiv.org/abs/hep-ph/0607121}{{\ttfamily
  hep-ph/0607121}}]
  [\href{http://inspirehep.net/search?p=find+Fleming:2006cd}{{\ttfamily
  InSPIRE}}].

\bibitem{Leibovich:2007vr}
A.~K. Leibovich and X.~Liu, {\it {The Color-singlet contribution to $e^{+}
  e^{-} \to J/\psi$ + $X$ at the endpoint}},
\href{http://dx.doi.org/10.1103/PhysRevD.76.034005}{{\em Phys.Rev.} {\bfseries
  D76} (2007) 034005} [\href{http://arxiv.org/abs/0705.3230}{{\ttfamily
  arXiv:0705.3230}}]
  [\href{http://inspirehep.net/search?p=find+Leibovich:2007vr}{{\ttfamily
  InSPIRE}}].

\bibitem{chenma2020}
A.-P. Chen and Y.-Q. Ma, {\it {In preparation}}, .

\bibitem{Lee:2007kg}
J.~Lee, H.-K. Noh, and C.-H. Yu, {\it {One-loop scalar integrals contributing
  to resummation of relativistic corrections to $\Gamma[J/\psi \to e^+ e^-]$}},
\href{http://dx.doi.org/10.3938/jkps.50.403}{{\em J. Korean Phys. Soc.}
  {\bfseries 50} (2007) 403--408}
  [\href{http://inspirehep.net/search?p=find+Lee:2007kg}{{\ttfamily InSPIRE}}].

\bibitem{Braaten:1995ej}
E.~Braaten and S.~Fleming, {\it {QCD radiative corrections to the leptonic
  decay rate of the B(c) meson}},
\href{http://dx.doi.org/10.1103/PhysRevD.52.181}{{\em Phys. Rev.} {\bfseries
  D52} (1995) 181--185} [\href{http://arxiv.org/abs/hep-ph/9501296}{{\ttfamily
  hep-ph/9501296}}]
  [\href{http://inspirehep.net/search?p=find+Braaten:1995ej}{{\ttfamily
  InSPIRE}}].

\end{thebibliography}
\providecommand{\href}[2]{#2}\begingroup\raggedright\endgroup


\end{document}